\documentclass[floatfix,aps,prd,superscriptaddress,twocolumn,preprintnumbers]{revtex4-1}

\usepackage{float}
\usepackage{amsmath,amssymb,bm,bbm}
\usepackage{graphicx,graphics,color}
\usepackage[colorlinks=true,linkcolor=blue,citecolor=blue,urlcolor=blue]{hyperref}
\usepackage{dsfont}
\usepackage{slashed}
\usepackage{mathtools}
\usepackage{feynmp-auto}
\usepackage{tikz}
\usepackage{tabularx}
\usepackage[utf8]{inputenc}
\usepackage{ulem}
\usepackage{multirow}
\usepackage{dcolumn}
\usepackage{soul}

\newcommand{\mycbox}[1]{\tikz{\path[draw=#1,fill=#1] (0,0) rectangle (.2cm,.2cm);}}

\begin{document}
\preprint{JLAB-THY-21-3539}

\title{Octet and decuplet baryon $\sigma$-terms and mass decompositions}

\author{P. M. Copeland}
\affiliation{\mbox{Department of Physics, Duke University, Durham, North Carolina 27708, USA}}
\author{Chueng-Ryong Ji}
\affiliation{\mbox{Department of Physics, North Carolina State University, Raleigh, North Carolina 27607, USA}}
\author{W. Melnitchouk}
\affiliation{Jefferson Lab, Newport News, Virginia 23606, USA}

\begin{abstract}
We present a comprehensive analysis of the SU(3) octet and decuplet baryon masses and $\sigma$-terms using high-precision lattice QCD data and chiral SU(3) effective theory with finite range regularization.
The effects of various systematic uncertainties, including from the scale setting of the lattice data and the regularization prescriptions, are quantified.
We find the pion-nucleon and strange nucleon $\sigma$-terms to be
    $\sigma_{\pi N} = 44(3)(3)$~MeV and
    $\sigma_{N s} = 50(6)(1)$~MeV, respectively.  
The results provide constraints on the energy-momentum tensor mass decompositions of the SU(3) octet and decuplet baryons, where we find the trace anomaly and quark/gluon energies decrease for strange baryons due to their larger strange $\sigma$-terms.
\end{abstract}

\date{\today} 
\maketitle

\section{Introduction}

With the scheduled construction of the Electron-Ion Collider and plans underway for elaborating its physics program, the origin of the nucleon's mass has been identified as one of the most important problems in hadronic physics for the next decade \cite{EIC:2021}.
In any decomposition of a baryon's mass, matrix elements that quantify its scalar quark content, usually referred to as $\sigma$-terms, play a crucial role \cite{Ji:1995prd, Ji:1995prl, Yang:2018, Liu:2021, Metz:2020, Lorce:2017, Ji:2021, Hatta:2018sqd, Tanaka:2018nae, Lorce:2021xku}. 
Additionally, $\sigma$-terms can be used to calculate the trace anomaly of the energy-momentum tensor (EMT), whose existence may explain quark confinement~\cite{Ji:1995prd, Liu:2021, Ji:2021pys}, and are important for understanding chiral symmetry breaking in quantum chromodynamics (QCD).
They also frequently appear in dark matter models in the couplings of spin-1/2 dark matter particles to scalar quark bilinears~\cite{Giedt:2009mr, Beane:2013, Hill:2015, Hoferichter:2019, Davoudi:2021, Batell:2019}. 
For these and other applications it is crucial to have precise and accurate determinations of baryon $\sigma$-terms.

The best determined $\sigma$-terms are those of the nucleon.
Its light ($u$ and $d$) quark contribution, commonly referred to as the pion-nucleon $\sigma$-term, $\sigma_{\pi N}$, was determined by Gasser {\it et al.}~\cite{Gasser:1990} from an analysis of $\pi N$ scattering in the early 1990s.
These studies, along with $\sigma$-term extractions from chiral extrapolations of baryon mass lattice QCD data~\cite{Aoki:2019, Young:2010, Ren:2012, Shanahan:2012wh, Shanahan:2013cd, Ren:2015}, predict a ``canonical'' magnitude of $\sigma_{\pi N} \approx 45$~MeV.
This value is also typically supported by direct lattice simulations of $\sigma_{\pi N}$~\cite{Aoki:2019, BMW:2016, RQCD:2016, ETM:2020}.
More recent calculations of $\sigma_{\pi N}$ determined from pionic atom scattering experiments predict a magnitude of around 60~MeV~\cite{Alarcon:2012, Hoferichter:2015, Friedman:2019, Alarcon:2021}, in tension with the smaller canonical predictions.
A flavor decomposition of the larger $\sigma_{\pi N}$ was made using baryon chiral perturbation theory in an effort to understand this discrepancy~\cite{Severt:2019}.

The strange nucleon $\sigma$-term, $\sigma_{Ns}$, is even more controversial. 
Chiral extrapolations of baryon mass data using the Feynman-Hellmann theorem have led to a wide range of values in the literature, from small negative to $\approx 100$~MeV~\cite{Aoki:2019, Young:2010, Ren:2012, Shanahan:2012wh, Shanahan:2013cd, Ren:2015}. 
Direct calculations of $\sigma_{Ns}$ on the lattice slightly narrow the estimates to $\sigma_{Ns} \approx$~20--100~MeV~\cite{Aoki:2019, BMW:2016, RQCD:2016, ETM:2020}.
However, disconnected diagrams and numerical derivatives about the physical point necessary for calculating $\sigma$-terms on the lattice are more computationally demanding than for baryon masses.
Improving chiral extrapolations of $\sigma$-terms from baryon mass data offers a cheaper route to computing other SU(3) octet and decuplet baryon $\sigma$-terms, in addition to the nucleon's $\sigma$-term.

Extracting $\sigma$-terms from lattice mass data does, on the other hand, require control of systematic uncertainties. 
Shanahan {\it et al.}~\cite{Shanahan:2012wh, Shanahan:2013cd} observed that the available baryon mass data can depend strongly on the lattice scale setting scheme, which can lead to strong variations in $\sigma_{Ns}$ without a proper continuum extrapolation.
Further uncertainties arise from the forms used for the chiral extrapolations of the lattice data to the physical point.

In particular, the convergence of the chiral expansion as a function of the meson mass can depend on the regularization scheme chosen, and it has been argued that finite range regularization (FRR) schemes, which involve an effective resummation of higher order terms in the baryon mass, parametrized by a finite range regulator, can provide better convergence over a larger range of masses~\cite{Young:2010, Young:2002ib, Donoghue:1998bs, Thomas:2002sj}.
To accurately capture the light and strange quark dependence of the baryon properties, it is important to use an effective field theory that can describe the strangeness content of baryons, such as SU(2) chiral perturbation theory for hyperons with the strange quark treated as heavy, or SU(3) chiral effective theory where the strange quark is treated as a light quark along with the $u$ and $d$ quarks.

In this paper, we employ the latest calculations of the octet and decuplet baryon masses within a relativistic chiral SU(3) effective theory framework~\cite{Copeland:2021} that uses FRR to analyze high-precision lattice QCD data from the PACS-CS~\cite{Aoki:2008sm} and QCDSF-UKQCD~\cite{Bietenholz:2011} Collaborations.
Using the Feynman-Hellmann theorem~\cite{Shanahan:2012wh, Young:2010}, we then use the parameters determined in the global fits to extract the light-quark and strange-quark $\sigma$-terms for all SU(3) octet and decuplet baryons, and discuss the impact of the results on baryon mass decompositions. 

In Sec.~II of this paper we begin with definitions of the baryon $\sigma$-terms and their chiral expansions, and extract the expansion parameters from lattice data in Sec.~III.
We discuss the resulting $\sigma$-terms in Sec.~IV and implications for the mass decompositions of SU(3) baryons in Sec.~V.
We conclude in Sec.~VI with some remarks about future extensions of this work, and in the Appendix present details of the results for the generalized mass expansion scheme.

\section{Baryon masses and $\sigma$-terms}
The $\sigma$-terms for a baryon ${\cal B}$ are defined as the forward baryon matrix elements of a quark scalar current of flavor~$q$,
\begin{equation}
    \sigma_{{\cal B} q} = m_q\, \langle {\cal B}|\overline{q}q |{\cal B} \rangle,
\qquad
    f_{{\cal B} q} = \frac{\sigma_{{\cal B} q}}{M_{\cal B}},
\label{eq: sigma term definition}
\end{equation}
where $m_q$ and $M_{\cal B}$ are the quark and baryon masses, respectively, and $f_{{\cal B}q}$ are the corresponding quark mass fractions.
For the nucleon, the $\pi N$ and strange $\sigma$-terms are defined as
  $\sigma_{\pi N} \equiv \sigma_{N \ell}
  = m_\ell\, \langle N | \overline{u} u + \overline{d}{d} | N \rangle$
and
  $\sigma_{Ns} = m_s\, \langle N | \overline{s}{s}| N \rangle$,
respectively, where the average light quark mass is $m_\ell = (m_u+m_d)/2$.
From the $m_q$ dependence of the baryon masses, one can compute $\sigma_{{\cal B}q}$ using the Feynman-Hellmann theorem,
\begin{equation}
\label{eq: Feynman-Hellmann}
    \sigma_{{\cal B} q} = m_q \frac{\partial M_{\cal B}}{\partial m_q}.
\end{equation}
In a chiral expansion of the baryon mass one can write~\cite{Shanahan:2012wh, Young:2010},
\begin{equation}
\label{eq: chiral mass expan}
    M_{\cal B}
    = M_{\cal B}^{(0)} + \delta M_{{\cal B}}^{(1)} + \delta M_{\cal B}^{(3/2)} + \cdots,
\end{equation}
where $M_{\cal B}^{(0)}$ is the baryon mass in the chiral limit, $m_q \to 0$, and $\delta M_{{\cal B}}^{(1)}$ and $\delta M_{\cal B}^{(3/2)}$ are quark mass dependent corrections.

The first correction is linear in the quark masses,
\begin{equation}
     \delta M_{{\cal B}}^{(1)} 
     = - C_{{\cal B}\ell}^{(1)} m_\ell - C_{{\cal B}s}^{(1)} m_s,
\label{eq: first order term} 
\end{equation}
with coefficients $C_{{\cal B}\ell}^{(1)}$ and $C_{{\cal B}s}^{(1)}$ determined from the chiral SU(3) effective theory~\cite{WalkerLoud:2004hf}.
These coefficients for octet (${\cal B}=B$) and decuplet (${\cal B}=T)$ baryons are linear combinations of the shared parameters, $\alpha$, $\beta$ and $\sigma$ for octet, and $\gamma$ and $\overline{\sigma}$ for decuplet,
\begin{subequations}
\begin{eqnarray}
C_{Bq}^{(1)}
&=& a_B^q\, \alpha
 +  b_B^q\, \beta
 +  c_B^q\, \sigma, 
\\
C_{Tq}^{(1)}
&=& a_T^q\, \gamma\,
+\, b_T^q\, \bar\sigma,
\end{eqnarray}
\end{subequations}
where the values of the constants $a_{\cal B}^q$, $b_{\cal B}^q$, and $c_{\cal B}^q$ for $q=\ell, s$ are given in Ref.~\cite{WalkerLoud:2004hf}.

The $\delta M_{\cal B}^{(3/2)}$ term arises from the meson loop self-energies of the baryons, $\Sigma_{{\cal B B'} \phi}$, where ${\cal B'}$ and $\phi$ denote the intermediate baryon and meson in the loop. 
The unique feature of this correction is that it is nonanalytic in the quark mass $m_q \sim m_\phi^2$, according to the Gell-Mann--Oakes--Renner (GOR) relation~\cite{Gell-Mann:1968hlm}, with a low-energy structure that is model independent.
In this work we use the latest calculations for the baryon self-energies computed within a relativistic chiral SU(3) effective theory regularized with FRR~\cite{Copeland:2021}, which introduces an additional cutoff parameter $\Lambda_{\cal B}$.

To avoid mixing between the $\delta M_{\cal B}^{(3/2)}$ term and the lower order analytic $M_{\cal B}^{(0)}$ and $\delta M_{{\cal B}}^{(1)}$ terms, we ``renormalize'' the self-energies by subtracting the values of the ${\cal O}(m_\phi^0)$ and ${\cal O}(m_\phi^2)$ terms at $m_\phi=0$,
\begin{equation}
\label{eq:renorm}
    \overline{\Sigma}_{{\cal B B'}\phi}
    = \Sigma_{{\cal B B'}\phi}
    - \Sigma_{{\cal B B'}\phi}(0) 
    -  m_\phi^2 \frac{\partial \Sigma_{{\cal B B'}\phi} }
                     {\partial m_\phi^2}(0). 
\end{equation}
The explicit expressions for the self-energies $\Sigma_{{\cal B B'}\phi}$ are given in Ref.~\cite{Copeland:2021}.
This allows $\delta M_{\cal B}^{(3/2)}$ to be simply written as a sum of the renormalized self-energies over all ${\cal B'}$ and $\phi$ states,
\begin{equation}
     \delta M_{\cal B}^{(3/2)} = \sum_{{\cal B'} \phi} \overline{\Sigma}_{{\cal B B'} \phi}.
\label{eq: M three halves}
\end{equation}
To preserve SU(3) symmetry, we set all octet baryon masses in the self-energy equations to the average experimental octet mass, $M_B = 1142$~MeV, and all decuplet baryon masses to their experimental average, $M_T = 1455$~GeV, which gives an octet-decuplet mass difference of 313~MeV.

\section{Baryon mass parameters from lattice data}

Computing the ${\cal B} q$ $\sigma$-terms requires determining the parameters
    $\{M_{\cal B}^{(0)}, 
    \alpha, \beta, \sigma, \gamma, \overline{\sigma}, 
    \Lambda_{\cal B}\}$
in the various terms of Eq.~(\ref{eq: chiral mass expan}), which can be done by analysing lattice QCD data on octet and decuplet baryons as a function of quark mass.
Such data are available from the PACS-CS~\cite{Aoki:2008sm} and QCDSF-UKQCD~\cite{Bietenholz:2011} Collaborations, with the latter dataset for $N_f=2+1$ flavors particularly useful for studying variations with both $m_\ell$ and $m_s$.
More recent calculations at or near the physical point exist, but do not provide the light and strange quark mass dependence of octet and decuplet baryon masses required for our analysis.
We fit the lattice baryon mass data using Eq.~(\ref{eq: chiral mass expan}) as a function of both the $\pi$ and $K$ masses, using the GOR relation~\cite{Gell-Mann:1968hlm} to relate the quark and meson masses,
    $m_\ell \propto  m_\pi^2/2$
and
    $m_s \propto  m_K^2 - m_\pi^2/2$~\cite{Shanahan:2011, WalkerLoud:2004hf}.
For the $\eta$ meson loops in $\delta M_{\cal B}^{(3/2)}$, the $\eta$ mass can be obtained via
    $m_\eta^2 \propto \frac{2}{3}(m_\ell + 2m_s) 
              \to (4m_K^2 - m_\pi^2)/3$~\cite{WalkerLoud:2004hf}.

When converting data from lattice units to physical units, the method used to determine the lattice spacing~$a$ at each quark mass can have a significant impact on the magnitude of the strange quark $\sigma$-terms~\cite{Shanahan:2013cd}. 
A~standard practice for setting the scale is to assume that the lattice spacing remains fixed at each quark mass simulation point, which we refer to as the mass independent lattice spacing (MILS) scheme. 
In this case the lattice spacing is provided as $a = 0.0907$~fm for the PACS-CS points~\cite{Aoki:2008sm} and $a = 0.075$~fm for the QCDSF-UKQCD data~\cite{Bietenholz:2011}.
These values are also close to those determined self-consistently using chiral EFT~\cite{Lutz:2014}.

Alternatively, the scale can be chosen by relating it to an external quantity that is invariant under changes in the quark masses, so that any variation observed in a lattice simulation must be attributed to changes in the lattice spacing.
Referring to this as the mass dependent lattice spacing (MDLS) scheme, we choose the Sommer scale $r_0$ = 0.4921~fm for the PACS-CS data~\cite{Aoki:2008sm}, and the SU(3) singlet quantity
    $X_N = (M_N + M_\Sigma + M_\Xi)/3$ 
for the QCDSF-UKQCD data~\cite{Bietenholz:2011} to set the scale.

We emphasize that the choice of scale setting scheme is very relevant for the data sets used in this work, which do not have a proper continuum limit, and we cannot perform a robust scale setting analysis because of the limited range of coupling and quark mass parameter available.
Since the $\sigma$-terms are determined from derivatives of the baryon masses with respect to the quark mass [Eq.~(\ref{eq: Feynman-Hellmann})], they are very sensitive to any relative variations of the baryon mass data that come from changing the scale, and the lattice data should be handled with great care. 
This is well demonstrated by the large discrepancy found by Shanahan {\it et al}.~\cite{Shanahan:2013cd} for the strange nucleon $\sigma$-term, which varies  between $\sigma^{\rm MILS}_{Ns} = 59(6)$~MeV and $\sigma^{\rm MDLS}_{Ns} = 21(6)$~MeV for the mass independent and mass dependent schemes, respectively. 
On the other hand, we should also note that scale setting is not typically a concern for more modern lattice simulations which do have proper continuum extrapolations. 
A new lattice data set that studies the light and strange quark mass dependence of the full SU(3) octet and decuplet with such a continuum extrapolation would be ideal to minimize the scale setting uncertainty.

We apply small, finite volume corrections to both sets of data using the expressions derived in Refs.~\cite{Geng:2011, Beane:2011, Ishikawa:2009}.
To minimize the effects of finite lattice volume, only data from the largest ($32^3 \times 64$) lattice volumes are selected. 
Data at larger pion masses, $m_\pi^2 \gtrsim 0.25$~GeV$^2$, which are more susceptible to the choice of scale, are excluded.
With these cuts, fitting to the PACS-CS and QCDSF-UKQCD data gives agreement between the two scale setting schemes, reducing the systematic uncertainty in the strange $\sigma$-term, $\sigma_{Bs}$.

\begin{figure}[t]
\centering
\includegraphics[width=\linewidth]{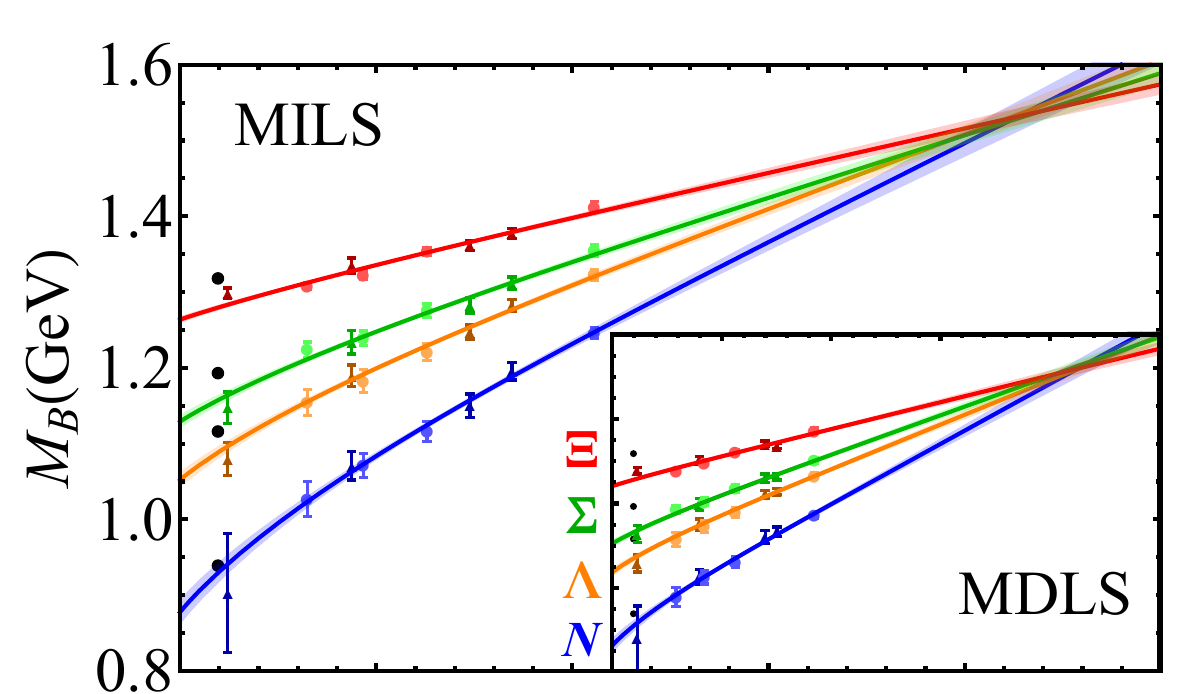}
\includegraphics[width=\linewidth]{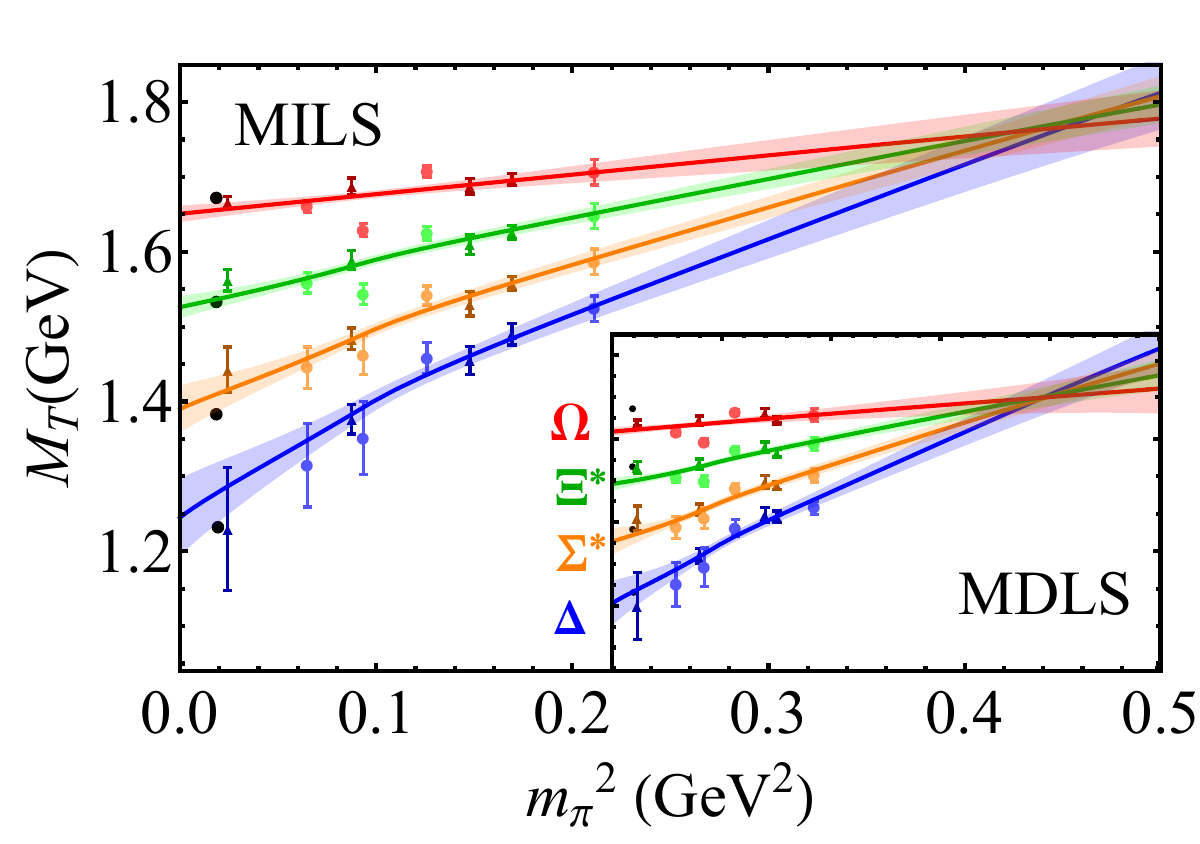}
\caption{Fits to the octet (top) and decuplet (bottom) baryon masses versus $m_\pi^2$ from the finite volume corrected \mbox{PACS-CS}~\cite{Aoki:2008sm} (lighter circles) and QCDSF-UKQCD~\cite{Bietenholz:2011} (darker triangles) data, using the MILS (main panels) and MDLS (insets) schemes. The experimental masses (black circles) are not used in the fit.}
\label{fig: Combined Hybrid fits}
\end{figure}

In fitting the free parameters we allow distinct chiral limit masses $M_B^{(0)}$ and $M_T^{(0)}$ for the octet and decuplet.
The parameters $\alpha$, $\beta$, $\sigma$, and $M_B^{(0)}$ are shared in the global fit of the octet, and similarly $\gamma$, $\overline{\sigma}$, and $M_T^{(0)}$ are shared in the fit of the decuplet.
In earlier work~\cite{Young:2010, Shanahan:2012wh} the cutoff $\Lambda_{\cal B}$ was taken as a shared parameter between all baryons.
In the present analysis, we allow distinct $\Lambda_{\cal B}$ values to be determined from fitting to the lattice data to parameterize the ultraviolet structure of each individual baryon ${\cal B}$ (although care should be taken when using SU(3) flavor-breaking regulators in calculations of other quantities).
We find, however, that this choice has a minimal impact on the results with slightly smaller uncertainties.

For the coupling constants, following Ref.~\cite{Copeland:2021} we use $D = 0.85$ and $F = 0.41$ for the octet-octet couplings, ${\cal C} = \frac65 g_A$ for the octet-decuplet, and ${\cal H} = \frac95 g_A$ for the decuplet-decuplet coupling, with $g_A = D + F = 1.26$ the axial vector charge.
For the pseudoscalar decay constant we use the convention where $f_\phi=93$~MeV.
The effects of varying the couplings within their uncertainties are relatively small \cite{Thomas:2013, Ren:2015}, and can be mostly compensated for by adjusting the free parameters.

The fits to the octet and decuplet baryon masses for the PACS-CS \cite{Aoki:2008sm} and QCDSF-UKQCD~\cite{Bietenholz:2011} data are presented in Fig.~\ref{fig: Combined Hybrid fits} for both the MILS and MDLS schemes.
The data include finite volume corrections, as well as corrections for the nonphysical values of the strange quark mass used in the simulations, although the latter does not affect the fit results.
Good agreement is obtained for the octet data, with $\chi^2$ per degree of freedom (dof) values $\chi^2_{\rm{dof}} = 0.78$ and 0.89 for the MILS and MDLS schemes, respectively.
The decuplet baryon data are more difficult to fit, with $\chi^2_{\rm{dof}}= 2.7$ and 3.3 for the MILS and MDLS cases, which is mostly due to the large spread of the lattice data and small uncertainties on the heavier baryon masses, most notably the $\Omega$ baryon. 
Improved calculations of decuplet baryons at small $m_\pi$ that take into account pion-nucleon scattering states may lead to improvements in the quality of the decuplet fits. 
Although the parameter values were found to be somewhat different for the MDLS and MILS fits, the extracted $\sigma$-terms turn out to be remarkably stable (see  Appendix~\ref{appendix} for information about the ``generalized" scheme, parameter values, and all $\sigma$-term results for the various schemes).

\begin{table*}[tb]
\centering
\caption{$\sigma$-terms ($\sigma_{{\cal B}q}$, $q=\ell, s$), baryon masses ($M_{\cal B}$), and their ratios ($f_{{\cal B}q}$), together with the trace anomaly ($f_{{\cal B}a}$) and the sum of the quark and gluon energy contributions $\langle x \rangle_{{\cal B}q}^E + \frac34 \langle x \rangle_{{\cal B}g}$, extracted from fits to lattice QCD data.
The first uncertainty is statistical, the second is systematic from the differences between the MILS and MDLS results, and the third on the $\sigma$-terms is from the theoretical uncertainty of higher order corrections in chiral effective theory.}
\begin{tabular}{l | c  r  r | l  l  l  l}
\hline
~${\cal B}$ &
~~$\sigma_{{\cal B}\ell}$~(MeV)~ & ~~$\sigma_{{\cal B}s}$~(MeV)~~ &
~~~$M_{\cal B}$~(MeV)~~~  &
~~~~~$f_{B\ell}$ &   ~~~~~~$f_{Bs}$ &  ~~~~~~$f_{{\cal B} a}$ &
$\langle x \rangle_{{\cal B}q}^E + \frac34 \langle x \rangle_{{\cal B}g}$
\\
\hline
~$N$
& ~44(3)(3)(4)    & ~50(6)(1)(10)~~~  & 920(10)(10)~~\!
& ~0.047(3)(3)(5)  & ~~0.053(6)(1)(10)~  & ~~0.900(7)(3)(11)   & ~~0.675(7)(3)(11)
\\
~$\Lambda$
& ~31(1)(2)(3)     & ~196(5)(7)(39)~~~ & 1080(6)(10)~~~
& ~0.028(1)(2)(3)  & ~~0.176(4)(6)(35)   & ~~0.796(4)(6)(35)   & ~~0.597(4)(7)(35)
\\
~$\Sigma$
& ~25(1)(1)(3)     & ~256(5)(7)(51)~~~  & 1145(5)(13)~~~ 
& ~0.021(1)(1)(2)  & ~~0.215(4)(6)(43)   & ~~0.764(4)(6)(43)   & ~~0.573(4)(6)(43)
\\
~$\Xi$  
& ~15(1)(1)(2)     & ~~365(5)(12)(73)~\, & 1269(3)(12)~~~  
& ~0.011(1)(1)(1)   & ~~0.277(4)(10)(55)  & ~~0.712(4)(10)(55)  & ~~0.534(4)(10)(55)
\\
& & & & & & &
\\
~$\Delta$ 
& ~29(9)(3)(3)     & ~67(11)(3)(13)~~\! & 1263(28)(23)~~\! 
& ~0.024(9)(2)(2)   & ~~0.054(9)(2)(10)   & ~~0.921(13)(3)(10)  & ~~0.692(13)(3)(10)
\\
~$\Sigma^*$ 
& ~18(6)(2)(2)     & ~189(11)(9)(38)~\,  & 1385(13)(22)~~\!
& ~0.013(4)(1)(1)   & ~~0.137(8)(7)(27)   & ~~0.850(9)(7)(27)   & ~~0.638(9)(7)(27)
\\
~$\Xi^*$
& ~10(3)(2)(1)     & ~307(12)(15)(61)  & 1520(6)(21)~~~
& ~0.007(2)(1)(1)   & ~~0.200(8)(10)(40)  & ~~0.793(8)(10)(40)  & ~~0.594(8)(10)(40)
\\
~$\Omega$ 
& ~~\,5(1)(1)(1)   & ~418(14)(20)(84)  & 1663(8)(18)~~~
& ~0.003(1)(1)(0)   & ~~0.250(8)(12)(50)  & ~~0.747(8)(12)(50)  & ~~0.560(8)(12)(50)
\\
\hline
\end{tabular}
\label{tab: sigmaterms}
\end{table*}

\section{Extracted $\sigma$-terms}

From the parameters determined through the fits to the baryon mass data, direct predictions can be obtained for the light and strange quark $\sigma$-terms. 
For the nucleon, we find excellent agreement of the results from the different scale setting prescriptions, with
    $\sigma_{\pi N} = 46(3)$~MeV and 41(4)~MeV,
and
    $\sigma_{Ns} = 49(8)$~MeV and 50(8)~MeV, 
for the MILS and MDLS schemes, respectively. 
This can be compared with the values 
    $\sigma_{Ns} = 59(6)$~MeV and 21(6)~MeV 
for the MILS and MDLS methods obtained by Shanahan {\it et al.} \cite{Shanahan:2012wh, Shanahan:2013cd}, who do not fit explicitly to the QCDSF-UKQCD data.

Other chiral extrapolations \cite{Camalich:2010, Ren:2015} have extracted $\sigma$-terms from lattice data using a covariant SU(3) baryon chiral perturbation theory with dimensional regularization. 
Camalich {\it et al.} \cite{Camalich:2010} fit to the PACS-CS data \cite{Aoki:2008sm} up to next-to-leading order, finding nucleon $\sigma$-terms $\sigma_{\pi N} = 59(2)(17)$~MeV and $\sigma_{N s} = -4(23)(25)$~MeV. 
Ren {\it et al.} \cite{Ren:2015} use a similar scheme, but up to N$^3$LO, fitting to the PACS-CS \cite{Aoki:2008sm}, QCDSF-UKQCD \cite{Bietenholz:2011}, and LHPC \cite{Walker-Loud:2009} data to obtain $\sigma_{\pi N} = 55(1)(4)$~MeV and $\sigma_{N s} = 27(27)(4)$~MeV.
Our results are somewhat different from both of these analyses due to the FRR scheme used, which offers better convergence of the mass expansions~\cite{Young:2002ib} (and therefore $\sigma$-term expansions) and produces smaller fit uncertainties.
Our fits also do not include experimental masses, whose small uncertainties could skew the slopes of the fitted masses.
This may explain our somewhat larger $\sigma_{Ns}$ compared with that in Ref.~\cite{Ren:2015}.

Our results can also be compared with chiral extrapolations of $\sigma$-terms from the LHPC lattice data \cite{Walker-Loud:2009}, which yield $\sigma_{\pi N}$ values of 84(17)(20)~MeV and 42(14)(9)~MeV using next-to-next-to-leading order heavy baryon chiral perturbation theory with and without the $\Delta$ resonance, respectively.
The differences with our results could be due to numerous reasons, the most obvious being the different regularizations and the masses used in the chiral extrapolation. 
Young and Thomas \cite{Young:2010} performed more sophisticated extrapolations by collectively fitting the LHPC data with the PACS-CS data using heavy baryon chiral perturbation theory with FRR, finding results consistent with Refs.~\cite{Shanahan:2012wh, Shanahan:2013cd} and differing with our findings for presumably the same reasons.

We also note that the QCDSF-UKQCD collaboration has performed chiral extrapolations of the full octet baryons using their own data \cite{QCDSF:Hyperon}. 
Their results are mostly consistent with ours within uncertainties, with the exception of the somewhat smaller light nucleon $\sigma_{Nl} = 31(3)(4)$.
This could be due to several reasons, such as not using chiral perturbation theory to extrapolate the $\sigma$-terms~\cite{QCDSF:Hyperon}, but rather 
expanding about the SU(3) flavor line, as well as the inclusion of the PACS-CS data in our analysis.

The agreement between our MILS and MDLS results persists for all other baryons in the SU(3) octet and decuplet. 
In Table~\ref{tab: sigmaterms} we show the averaged values for the mass independent and mass dependent results, with differences between the two quoted as a systematic uncertainty.
To explore the model dependence of the results from the renormalization prescription and the imposition of SU(3) symmetry, we also perform a less restricted fit in which no parameters are shared between the baryons, apart from the chiral limit masses, $M^{(0)}_{\cal B}$, and which does not use Eq.~(\ref{eq:renorm}) for the self-energies.
This fit, which we refer to as the ``generalized'' scheme, gives results consistent with the more constrained fits described above.
In particular, for the nucleon $\pi N$ $\sigma$-term we find 
    $\sigma_{\pi N} = 47(4)$~MeV and 45(5)~MeV 
and for the strange $\sigma$-term 
    $\sigma_{Ns} = 59(14)$~MeV and 68(15)~MeV
for the MILS and MDLS scenarios, respectively, suggesting the model used for the extrapolation of the lattice data is not overreaching (see Appendix~\ref{appendix}).

There is also some additional theoretical uncertainty from higher order corrections in the chiral effective theory.
These come into play for both the calculation of the baryon masses in Eq.~(\ref{eq: chiral mass expan}) and in higher order corrections to the meson masses which would give corrections to the Feynman-Hellmann theorem.
We expect the corrections to the $\sigma$-terms arising from higher order terms in the baryon mass expansion to be well accounted for by our FRR prescription, as discussed in more detail in Appendix~\ref{sec: Convergence}. 
However, since SU(3) meson chiral perturbation theory typically has order 20\%--30\% corrections from next-to-leading order, their effect on the Feynman-Hellmann theorem may not be neglegible, especially when converting the derivative with respect to the strange quark mass. 
As a conservative estimate, we quote an additional 10\% theoretical uncertainty for the $\sigma_{{\cal B} l}$ values and a 20\% uncertainty for the $\sigma_{{\cal B} s}$ values arising from these corrections.

\begin{figure}[t]
\centering
\includegraphics[width=\linewidth]{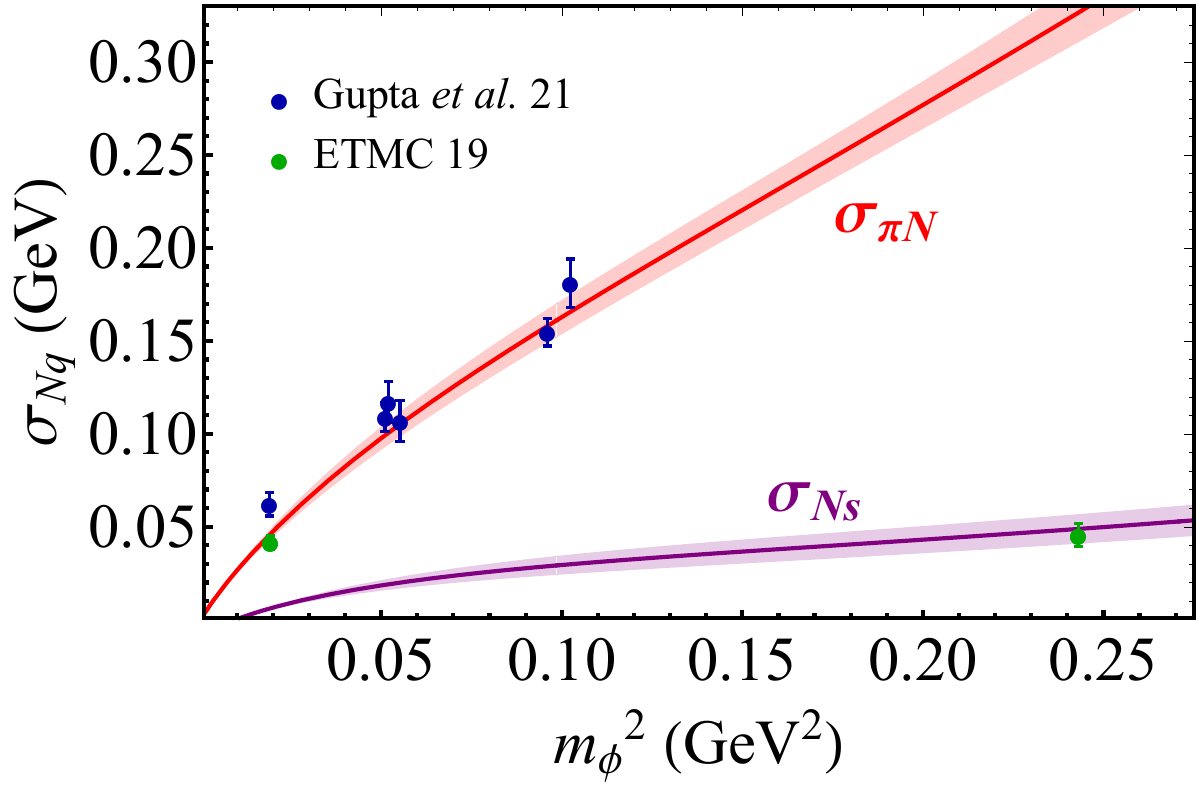}
\caption{Nucleon $\sigma$-terms compared with the Gupta {\it et al.}~\cite{Gupta:2021} (blue) and ETMC~\cite{ETM:2020} (green) direct calculations on the lattice. The $\sigma_{\pi N}$ is plotted as a function of $m_\pi^2$ with the strange quark mass fixed, and $\sigma_{Ns}$ as a function of $m_K^2$ with the pion mass fixed at the physical point.}
\label{fig: N sigma terms}
\end{figure}

Our results for $\sigma_{\pi N}$ and $\sigma_{N s}$ agree well with the averaged $N_f=2+1$ lattice values in the FLAG review~\cite{Aoki:2019}.
On the other hand, despite our comprehensive treatment of multiple data sets, scale setting schemes and fitting models, we cannot reconcile our results with those from the pionic atom and $\pi N$ scattering determinations of the $\pi N$ $\sigma$-term, $\sigma_{\pi N} \approx 60$~MeV~\cite{Alarcon:2012, Hoferichter:2015, Friedman:2019, Alarcon:2021}.
A~global analysis of experimental data and lattice simulations may be needed to understand this difference.
Recent results by Gupta {\it et al.}~\cite{Gupta:2021} have indicated that the discrepancy may be related to excited $\pi N$ and $\pi \pi N$ states in direct calculations of $\sigma_{\pi N}$ on the lattice. 
We note, however, that heavy baryon chiral perturbation theory is used for the computation of the scalar charge and $\sigma_{\pi N}$ in Ref.~\cite{Gupta:2021}, while a generalization to SU(3) is required for strange quark $\sigma$-terms, along with the other baryon $\sigma$-terms.
Additionally, the better convergence offered by FRR could provide an improved functional form to guide the lattice calculations. 
In Fig.~\ref{fig: N sigma terms} we compare our MILS $\sigma_{\pi N}$ results with direct calculations from Ref.~\cite{Gupta:2021, ETM:2020}, and the prediction for $\sigma_{N s}$ with Ref.~\cite{ETM:2020}.
The comparisons show good agreement for all masses, apart from the smallest $m_\pi^2$ results, which are of course the most important. 
This emphasizes the need for future simulations in order to resolve the discrepancy.

\section{Baryon mass decomposition}

The quark mass fractions $f_{{\cal B} q}$ in Eq.~(\ref{eq: sigma term definition}) represent the quark contributions to the baryon mass, which is defined by the matrix element of the EMT of QCD,
    $M_{\cal B} 
    = \langle {\cal B} | \int\!d^3 x\, T^{00}_{\mbox{\tiny \rm QCD}} | {\cal B} \rangle 
    / \langle {\cal B} | {\cal B} \rangle$. 
This can be decomposed according to \cite{Ji:1995prd,Ji:1995prl, Yang:2018, Liu:2021},
\begin{equation}
    M_{\cal B} 
    = \bigg[ 
        \sum_q \Big( \langle x \rangle_{{\cal B}q}^E 
                    + f_{{\cal B} q}
               \Big)
        + \frac34 \langle x \rangle_{{\cal B}g} 
        + \frac14 f_{{\cal B} a} 
      \bigg] M_{\cal B},
\label{eq: mass decomp}
\end{equation}
where 
    $\langle x \rangle_{{\cal B}q}^E
    = \frac34 
    \big( \langle x \rangle_{{\cal B}q} - f_{{\cal B} q}
    \big)$
is interpreted as the quark kinetic and potential energy, and
    $\langle x \rangle_{{\cal B}q,{\cal B}g}$
are the quark and gluon momentum fractions of the baryon at the scale $\mu$.
The trace anomaly of the EMT, $f_{{\cal B}a}$, can be computed from the sum rule
    $f_{{\cal B} a} + \sum_{q} f_{{\cal B} q} = 1$
\cite{Ji:1995prd, Ji:1995prl}.
Since the $\sigma$-term and trace anomaly contributions are scale independent, so is the sum, 
    $\sum_q \langle x \rangle_{{\cal B}q}^E + \frac34 \langle x \rangle_{{\cal B}g} = \tfrac34 f_{{\cal B} a}$. 
The numerical values of the various terms in Eq.~(\ref{eq: mass decomp}) are listed in Table~\ref{tab: sigmaterms}.

The decomposition (\ref{eq: mass decomp}) is scheme dependent and corresponds to defining the trace anomaly as the trace of the renormalized gluon component of the EMT.
An alternative decomposition defines the baryons mass from only the trace of the EMT ~\cite{Liu:2021, Metz:2020, Hatta:2018sqd, Tanaka:2018nae}.
Both of these decompostions are identical for octet and decuplet baryons, however, in a more general decomposition, such as those involving baryon gravitational form factors at zero momentum transfer, new form factors appear for the decuplet case.
For the temporal component of the EMT, these terms do not contribute due to the vanishing spin polarization for the decuplet for the $\mu=0$ component.
(For spatial components, they do not vanish, and can provide information on unique anisotropic terms in the pressure/work distributions of decuplet baryons, similar to those discussed for spin-1 hadrons \cite{Cosyn:2019}).

\begin{figure}[t]
\centering
\includegraphics[width=\linewidth]{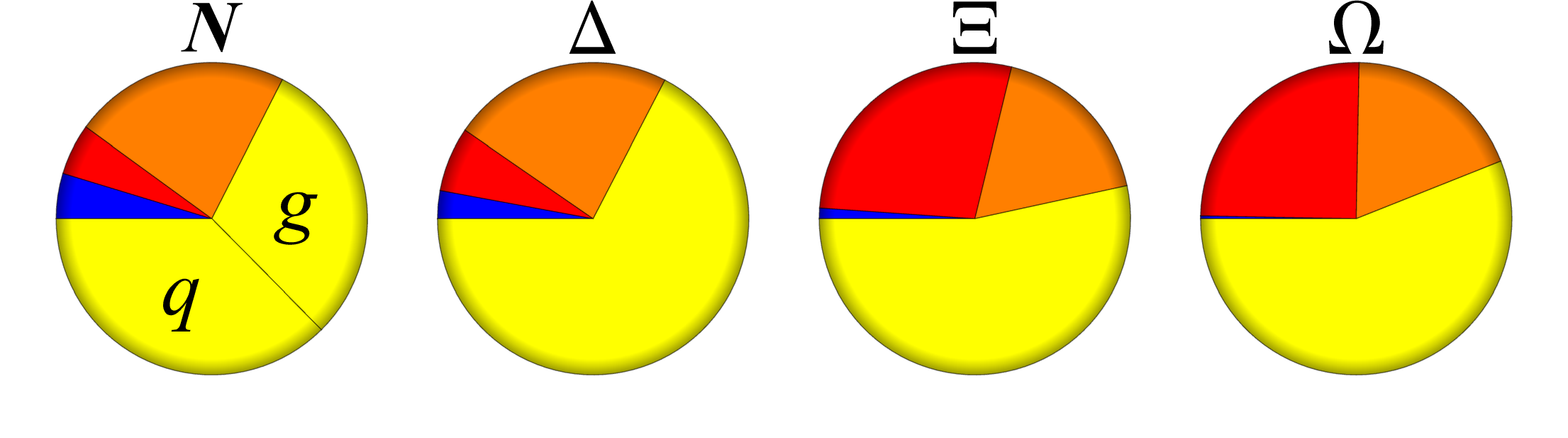}
\begin{tabular}{c c c c c c c c c c c}
    \mycbox{blue} & $f_{{\cal B}\ell}$ & & \mycbox{red} & $f_{{\cal B}s}$ & & \mycbox{orange} & $\frac14 f_{{\cal B}a}$ & & \mycbox{yellow} & $\langle x \rangle_{{\cal B}q}^E + \frac34 \langle x \rangle_{{\cal B}g}$
\end{tabular}
\caption{Mass decomposition of the nucleon, $\Delta$, $\Xi$, and $\Omega$ baryons in the rest frame, showing the fractional contributions of the light (blue) and strange (red) $\sigma$-terms, the trace anomaly (orange), and the sum of the quark and gluon energies (yellow). 
For the nucleon, the quark ($q$) and 3/4 the gluon ($g$) energies are shown separately, computed using PDFs from the JAM global QCD analysis~\cite{Cocuzza:2021cbi, Cocuzza:2021rfn} at $\mu = 2$~GeV.}
\label{fig: mass decomp}
\end{figure}

While the debate about the most appropriate mass decomposition continues~\cite{Liu:2021, Metz:2020, Lorce:2017, Ji:2021, Lorce:2021xku}, we note that the $\sigma$-term contribution to $M_{\cal B}$ is independent of the scheme.
In Fig.~\ref{fig: mass decomp} we use Eq. (\ref{eq: mass decomp}) to illustrate the decomposition for several representative octet and decuplet masses (nucleon, $\Delta$, $\Xi$ and $\Omega$) into their trace anomaly, quark energy, and gluon momentum components.

The latter contributions decrease with increasing magnitude of the quark mass fractions for heavier baryons, so that the sum of the quark mass fractions is $\approx$ 3 times larger for the $\Xi$ compared to the nucleon, for example.
This is of particular interest as the gluonic energy from the trace anomaly may be associated with quark confinement in hadrons (see {\it e.g.}, Refs.~\cite{Ji:1995prd, Liu:2021, Ji:2021pys}) by exerting a restoring pressure on the hadrons~\cite{Liu:2021, Ji:2021pys}, reminiscent of that in a bag model~\cite{Ji:1995prd}.
The decreasing magnitude of the trace anomalies for heavier baryons may indicate a proportionally smaller restoring pressure.

\section{Outlook}

It is clear from this analysis that to better understand the internal quark and gluon compositions of baryons in QCD, the quark and gluon momentum distributions of all octet and decuplet baryons need to be further studied using lattice and effective field theory techniques.
Experimentally, the nucleon mass decomposition will be one of the issues that will be probed at the future Electron-Ion Collider \cite{EIC:2021}, while the J-PARC facility in Japan will study the origin of hyperon masses \cite{Ohnishi:2019cif}.
Other applications of our results include constraining dark matter models, such as those involving WIMPs interacting with heavy nuclei \cite{Giedt:2009mr, Beane:2013, Hill:2015, Hoferichter:2019, Davoudi:2021}, which require nuclear $\sigma$-terms.
Our results for the light and strange quark $\sigma$-terms of octet and decuplet baryons with reduced systematic uncertainty can serve as a good basis for such endeavors.

\acknowledgments

We thank P.~E.~Shanahan, A.~W.~Thomas, and R.~D.~Young for helpful discussions. This work was supported by the U.S. Department of Energy Contract No. DE-AC05-06OR23177, under which Jefferson Science Associates, LLC operates Jefferson Lab and DOE Contract No. DEFG02-03ER41260.

\clearpage
\pagebreak
\pagebreak
\onecolumngrid
\appendix

\section{Generalized mass expansion scheme}
\label{appendix}
\vspace*{-0.2cm}

To explore the model dependence of our analysis of the baryon masses, we consider a generalized parametrization for $M_{\cal B}$, which does not impose the constraints from SU(3) symmetry and renormalization that were used in Eqs.~(\ref{eq: first order term})--(\ref{eq: M three halves}).
In this alternative scenario the coefficients
    $C_{{\cal B}\ell}^{(1)}$ and 
    $C_{{\cal B}s}^{(1)}$ 
in Eq.~(\ref{eq: first order term}) and $\Lambda_{\cal B}$ in $\delta M_{\cal B}^{(3/2)}$ are treated as free parameters to be determined for each baryon ${\cal B}$ from the data.
The only constraint imposed is that the chiral limit mass is the same for all baryons.
Additionally, in contrast to Eq.~(\ref{eq: M three halves}), the $\delta M_{\cal B}^{(3/2)}$ term is not renormalized and is given by the direct sum over all possible intermediate states for the self-energies,
\begin{equation}
    \delta M_{\cal B}^{(3/2)} 
    = \sum_{{\cal B'} \phi} \Sigma_{{\cal B B'} \phi}. 
\end{equation}
Since the parameters in this scheme are uncorrelated, in each self-energy term $\Sigma_{{\cal B B'} \phi}$ we use the appropriate physical baryon mass instead of the mass averaged over the multiplet.

\vspace*{-0.2cm}
\subsection{Fits to data}
\vspace*{-0.2cm}

\begin{figure}[H]
\centering
\includegraphics[width=0.42\linewidth]{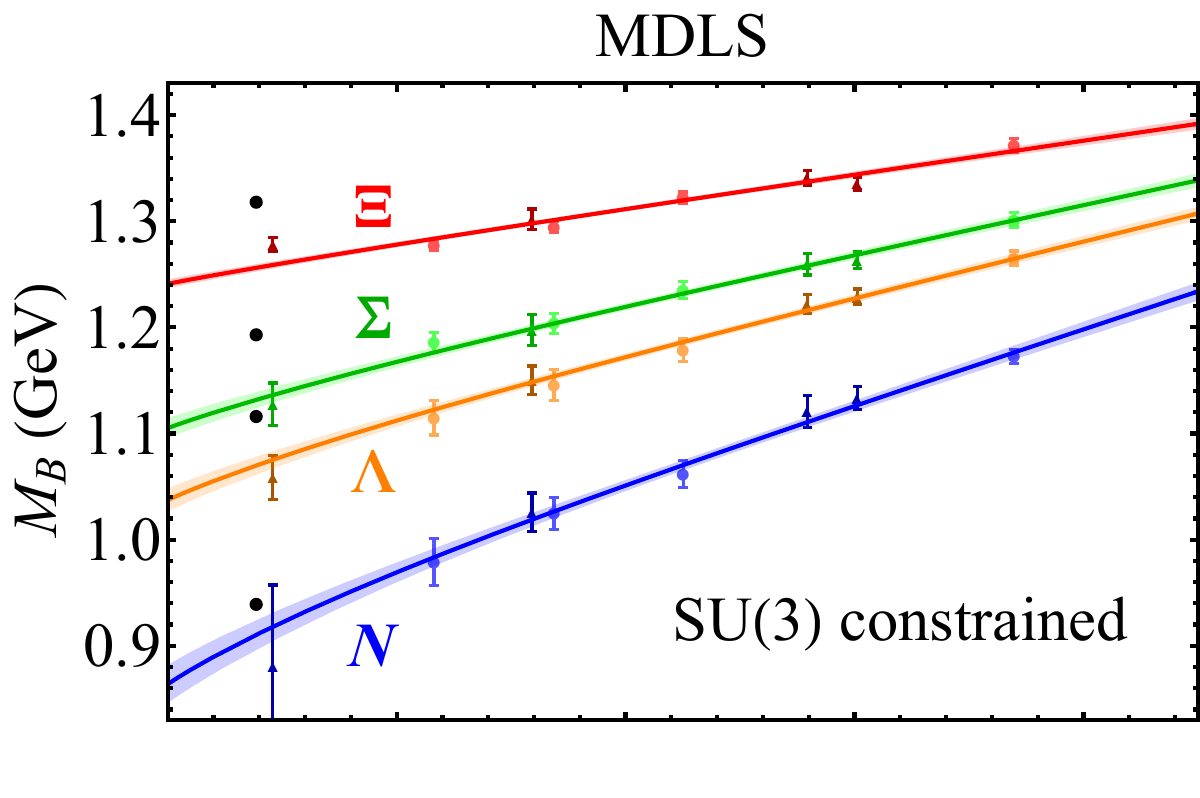}\hspace{-0.9cm}
\includegraphics[width=0.42\linewidth]{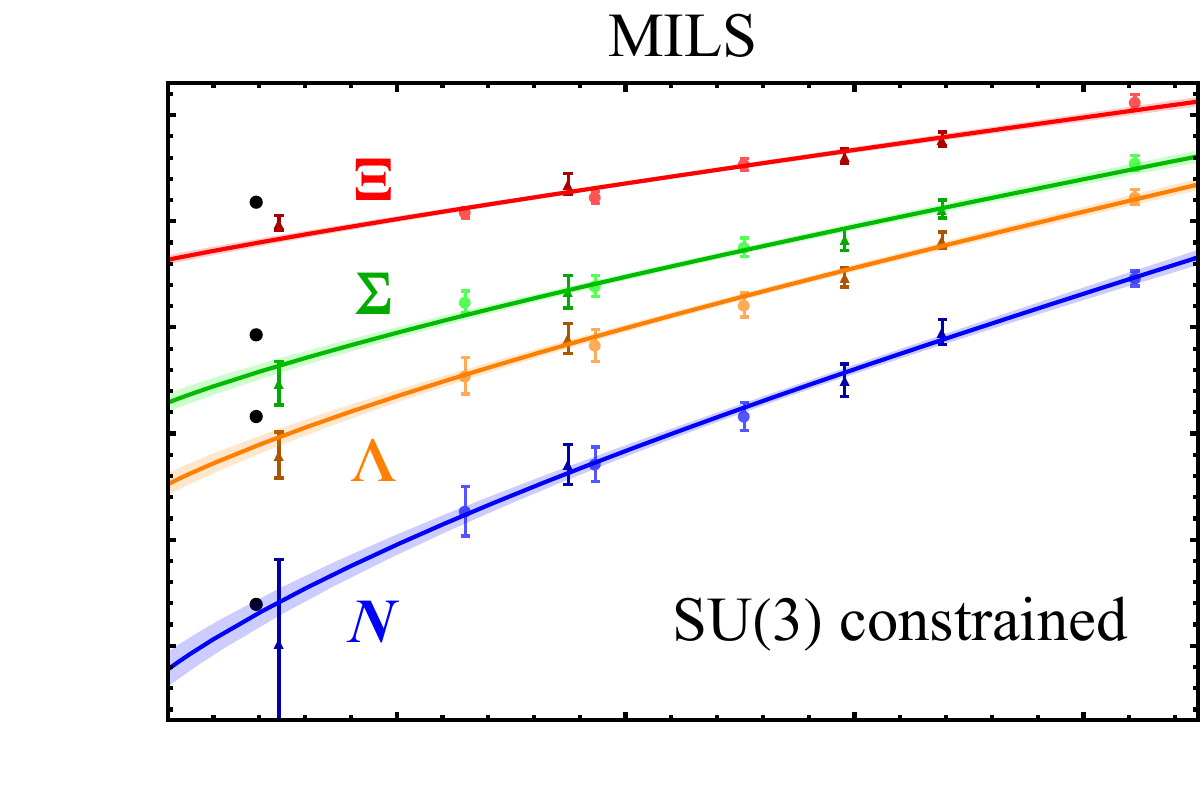}\vspace*{-0.3cm}
\includegraphics[width=0.42\linewidth]{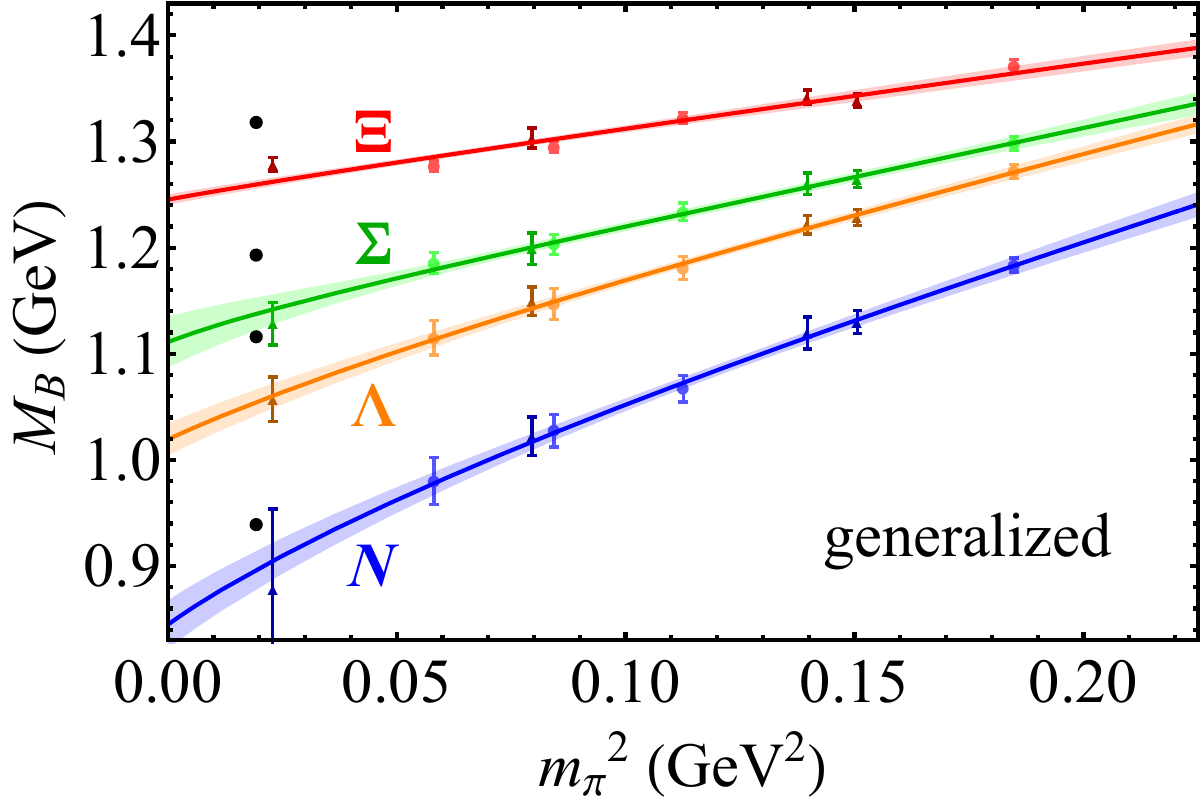}\hspace{-0.9cm}
\includegraphics[width=0.42\linewidth]{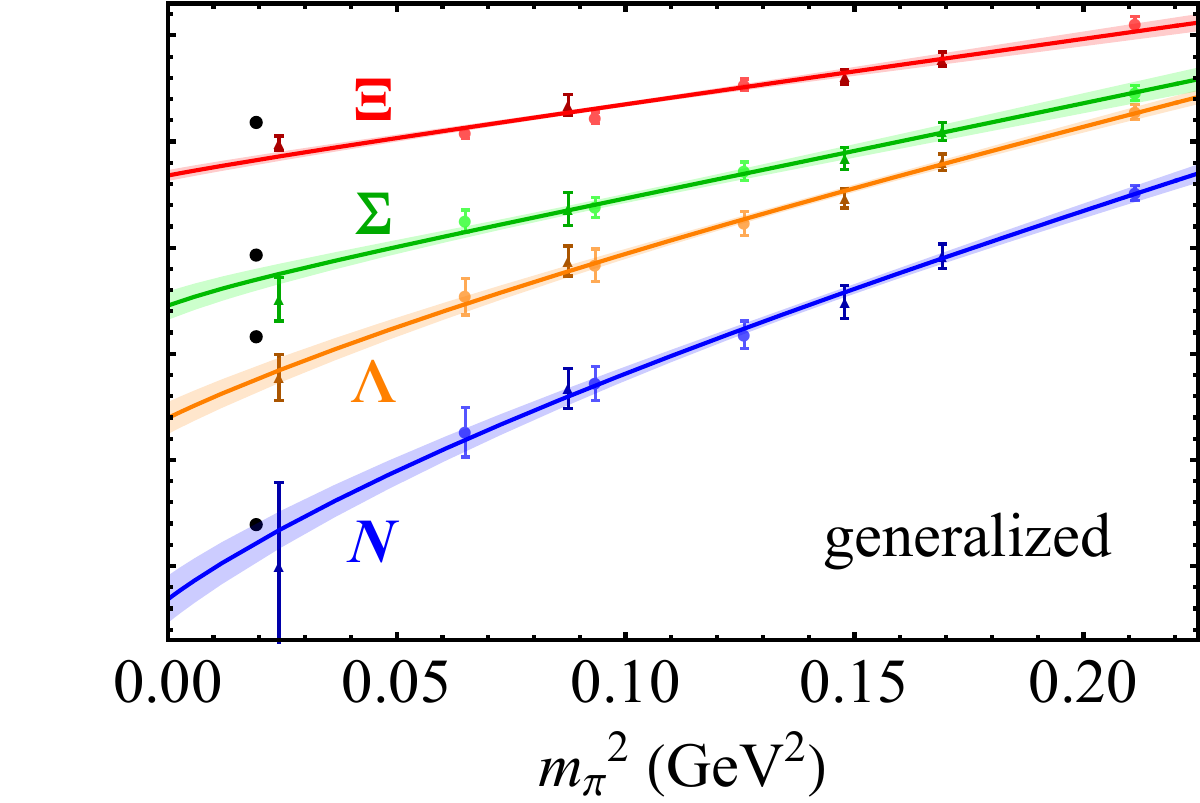}
\caption{Octet baryon masses versus $m_\pi^2$ for the $N$ (blue), $\Lambda$ (orange), $\Sigma$ (green) and $\Xi$ (red) baryons, for the MDLS (left) and MILS (right) schemes, with the standard SU(3) constrained (top) and generalized (bottom) mass expansion schemes. The fits are compared with PACS-CS (darker)~\cite{Aoki:2008sm} and QCDSF-UKQCD (lighter)~\cite{Bietenholz:2011} data, with the empirical values indicated by the black circles at the physical point.}
\label{fig: Combined Hybrid Octet fits}
\end{figure}

\begin{figure}[H]
\centering
\includegraphics[width=0.42\linewidth]{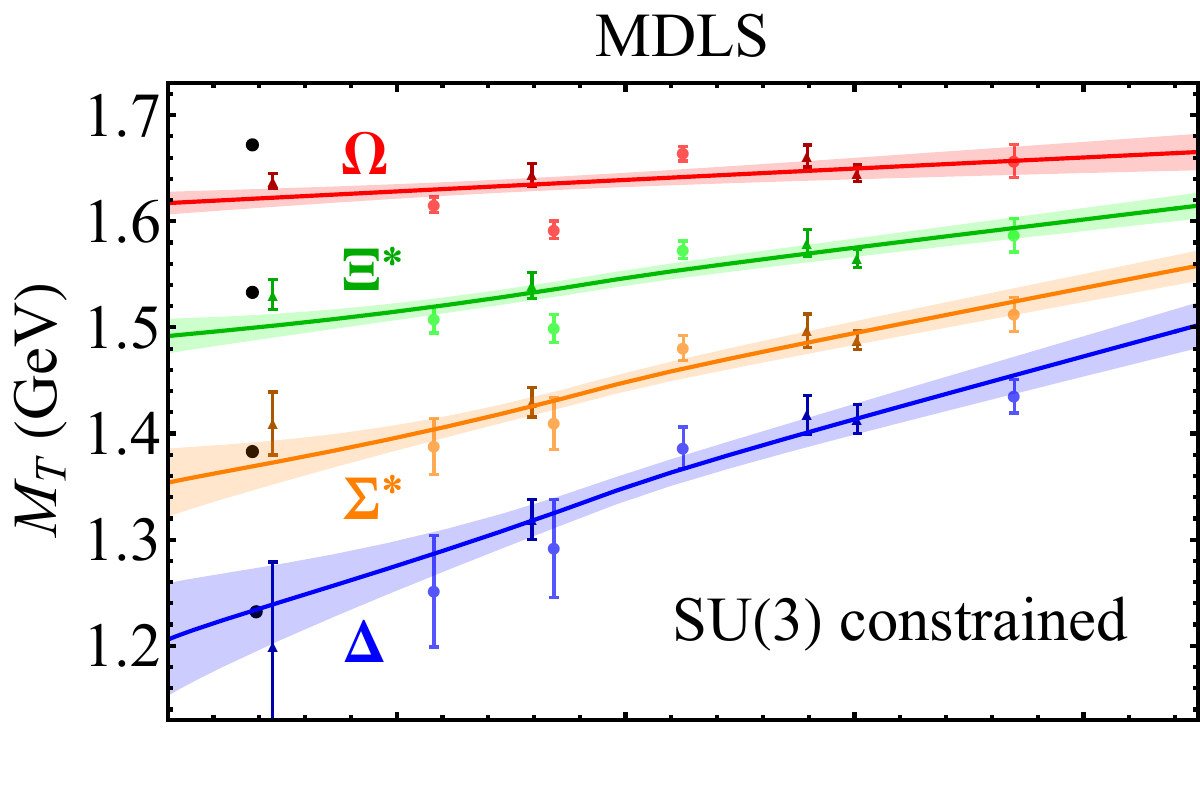}\hspace{-0.9cm}
\includegraphics[width=0.42\linewidth]{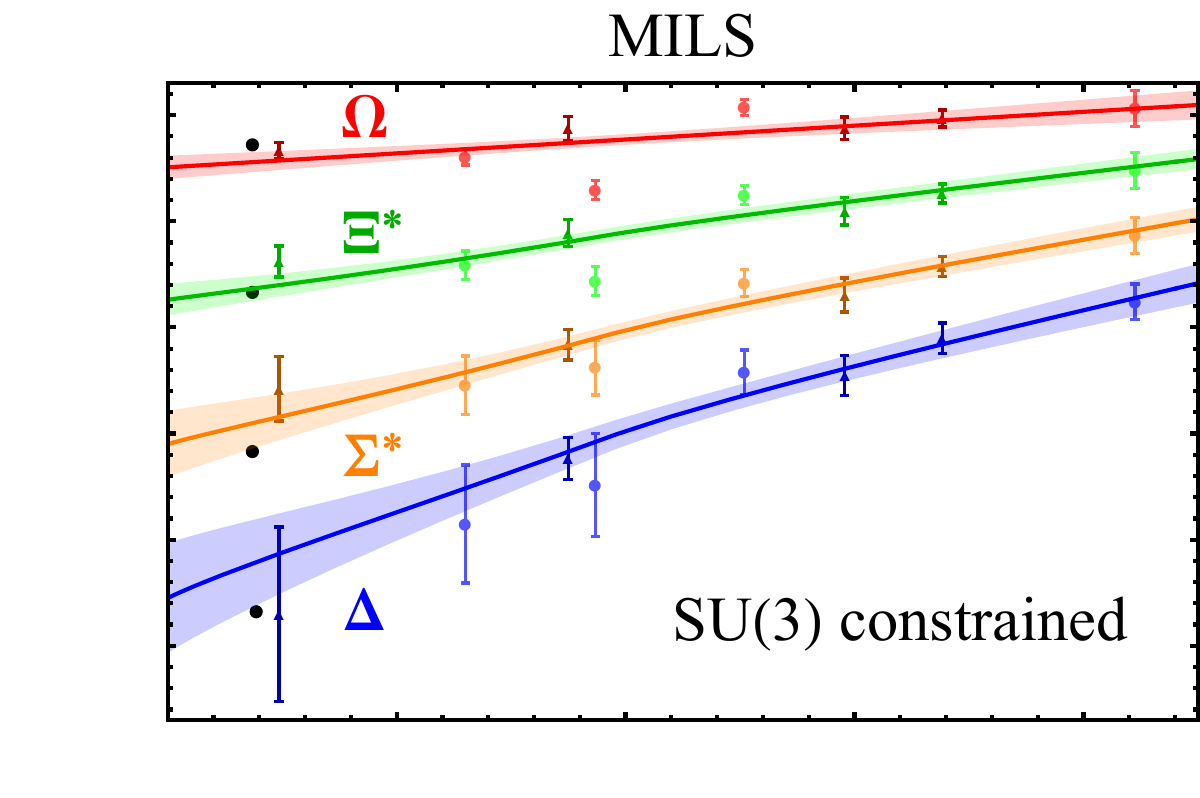}\vspace*{-0.3cm}
\includegraphics[width=0.42\linewidth]{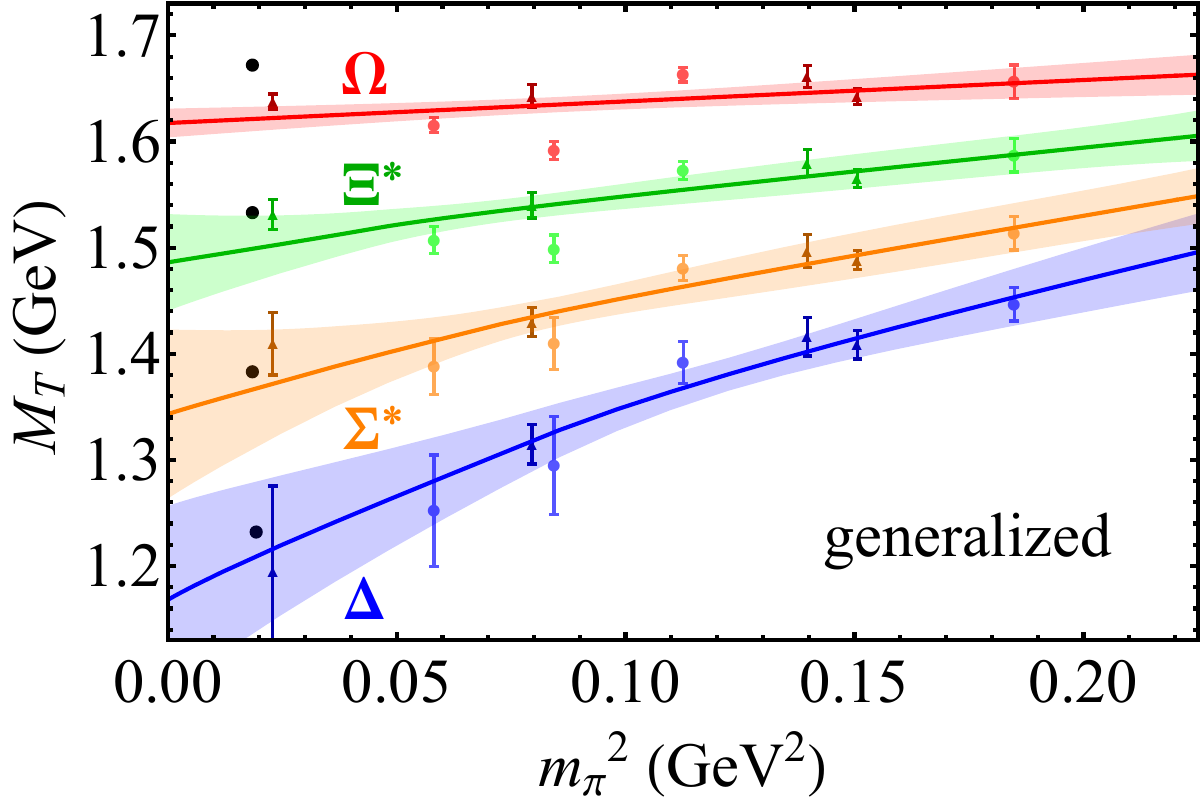}\hspace{-0.9cm}
\includegraphics[width=0.42\linewidth]{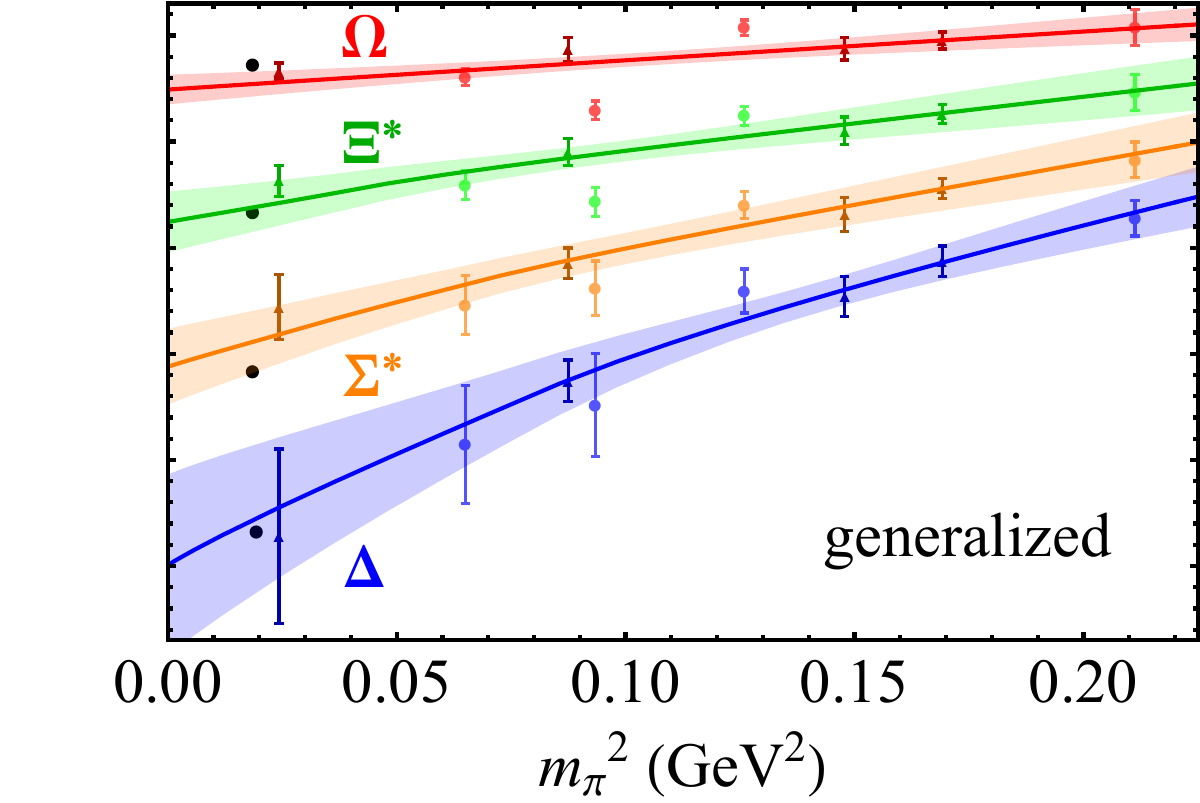}
\caption{As in Fig.~\ref{fig: Combined Hybrid Octet fits}, but for the decuplet baryons $\Delta$ (blue), $\Sigma^*$ (orange), $\Xi^*$ (green) and $\Omega$ (red).}
\label{fig: Combined Hybrid Decuplet fits}
\end{figure}

The results for the octet and decuplet baryon masses are given in Figs.~\ref{fig: Combined Hybrid Octet fits} and \ref{fig: Combined Hybrid Decuplet fits}, respectively, for the MDLS and MILS schemes, using the SU(3) constrained and generalized mass expansion schemes.
Note that the SU(3) constrained results are identical to those shown in Fig.~\ref{fig: Combined Hybrid fits}, which we include here for more direct comparison with the generalized results at small $m_\pi^2$ values.

\vspace*{-0.2cm}
\subsection{Fit parameters}
\vspace*{-0.2cm}
\label{WL expansion}

The relations for the $C_{{\cal B}\ell}^{(1)}$ and $C_{{\cal B}s}^{(1)}$ coefficients in the linear quark mass term, $\delta M_{\cal B}^{(1)}$, in terms of the fit parameters $\alpha$, $\beta$ and $\sigma$ for the octet, and $\gamma$ and $\overline{\sigma}$ for the decuplet, in the SU(3) constrained scenario are given in Table~\ref{table: WL octet} \cite{WalkerLoud:2004hf}.
\begin{table}[b]
\caption{Relations for the coefficients of the linear light quark, $C_{{\cal B}\ell}^{(1)}$, and strange quark, $C_{{\cal B}s}^{(1)}$, mass terms in terms of the fit parameters $\alpha$, $\beta$, $\sigma$ for the octet baryons and $\gamma$ and $\overline{\sigma}$ for the decuplet baryons \cite{WalkerLoud:2004hf}.\\}
\label{table: WL octet}
\centering
\begin{tabular}{ l | c c }
\hline
~${\cal B}$~& $C_{{\cal B}\ell}^{(1)}$  
            & $C_{{\cal B}s}^{(1)}$
\\
\hline
~$N$        & ~~$2\alpha + 2\beta + 4\sigma$~~
            & ~~$2\sigma$~~
\\ & &
\\
~$\Lambda$  & ~~~$\alpha + 2\beta + 4\sigma$~~   
            & ~~$\alpha + 2\sigma$~~
\\ & &
\\
~$\Sigma$   & ~~$\frac53 \alpha + \frac23 \beta + 4\sigma$~~   
            & ~~$\frac13 \alpha + \frac43 \beta + 2\sigma$~~
\\ & &
\\
~$\Xi$      & ~~$\frac13 \alpha + \frac43 \beta + 4\sigma$~~
            & ~~$\frac53 \alpha + \frac23 \beta + 2\sigma$~~
\\ & & 
\\
~$\Delta$   & ~~$2\gamma - 4\overline{\sigma}$~~
            & ~~$2\overline{\sigma}$~~ 
\\ & &
\\
~$\Sigma^*$ & ~~$\frac43 (\gamma-3\overline{\sigma})$~~   
            & ~~$\frac23 (\gamma-3\overline{\sigma})$~~
\\ & &
\\
~$\Xi^*$    & ~~$\frac23 (\gamma-6\overline{\sigma})$~~   
            & ~~$\frac23 (2\gamma-3\overline{\sigma})$~~
\\ & &
\\
~$\Omega$   & ~~$2\gamma - 2\overline{\sigma}$~~  
            & ~~$4\overline{\sigma}$~~
\\
\hline
\end{tabular}
\end{table}
The values of the octet parameters 
    \{$M_B^{(0)}$, $\alpha$, $\beta$, $\sigma$, $\Lambda_B$\} 
and decuplet parameters 
    \{$M_T^{(0)}$, $\gamma$, $\overline{\sigma}$, $\Lambda_T$\}
determined from the fits of the SU(3) constrained scheme are given in Tables~\ref{table: WL parameter values} and \ref{table: dec WL parameter values}, respectively. 
\begin{table}[t]
\begin{center}
\caption{Octet baryon fit parameters $M_B^{(0)}$, $\alpha$, $\beta$, $\sigma$, and the regulators $\Lambda_N$, $\Lambda_\Lambda$, $\Lambda_\Sigma$, and $\Lambda_\Xi$, for the SU(3) constrained mass expansion scheme and for the MDLS and MILS scenarios.  Statistical uncertainties are given in parentheses.}
\begin{tabular}{ l | c c c c | c c c c }
\hline
& $M_B^{(0)}$ 
& $\alpha$ & $\beta$ & $\sigma$ 
& $\Lambda_N$ & $\Lambda_{\Lambda}$ & $\Lambda_{\Sigma}$ & $\Lambda_{\Xi}$
\\
& (MeV) & (MeV$^{-1}$) & (MeV$^{-1}$) & (MeV$^{-1}$)
& (MeV) & (MeV) & (MeV) & (MeV)
\\
\hline
~{\bf MDLS}~ 
& ~799(27)~ & ~$-1\,362(100)$~ & $-1\,097(85)$~ & ~$-433(53)$~ 
& ~~535(81)~~ & ~~545(83)~~ & ~~544(78)~~ & ~~555(83)~~
\\
~{\bf MILS}~ 
& ~794(26)~ & $-1\,471(71)$~ & $-1\,241(63)$~ & $-513(43)$
& ~~687(70)~~ & ~~695(70)~~ & ~~703(66)~~ & ~~708(70)~~
\\
\hline
\end{tabular}
\label{table: WL parameter values}
\end{center}
\end{table}
\begin{table}[t]
\begin{center}
\caption{Decuplet baryon fit parameters $M_T^{(0)}$, $\gamma$, $\overline{\sigma}$, and the regulators $\Lambda_\Delta$, $\Lambda_{\Sigma^*}$, $\Lambda_{\Xi^*}$, and $\Lambda_\Omega$, for the SU(3) constrained mass expansion scheme and for the MDLS and MILS scenarios.  Statistical errors are given in parentheses.}
\begin{tabular}{ l | c c c | c c c c }
\hline
& $M_T^{(0)}$
& $\overline{\sigma}$
& $\gamma$
& $\Lambda_{\Delta}$ & $\Lambda_{\Sigma^*}$ & $\Lambda_{\Xi^*}$ & $\Lambda_{\Omega}$ 
\\
& (MeV) & (MeV$^{-1}$) & (MeV$^{-1}$)
& (MeV) & (MeV) & (MeV) & (MeV)
\\
\hline
~{\bf MDLS}~ 
& ~$1\,122(93)$ & ~$252(366)$~ & ~$-1\,377(467)$~
& ~$483(89)$~   & ~$487(91)$~  & ~$490(94)$~  & ~$492(100)$~
\\
~{\bf MILS}~ 
& ~$1\,136(91)$ & ~$462(267)$~ & ~$-1\,500(364)$~ 
& ~~\,$549(101)$~ & ~~\,$548(100)$~ & ~~\,$549(103)$~ & ~$552(107)$~
\\
\hline
\end{tabular}
\label{table: dec WL parameter values}
\end{center}
\end{table}
Similarly, the octet baryon parameters
    \{$M_B^{(0)}$, $C_{B\ell}^{(1)}$, $C_{Bs}^{(1)}$, $\Lambda_B$\} 
and decuplet baryon parameters
    \{$M_T^{(0)}$, $C_{T\ell}^{(1)}$, $C_{Ts}^{(1)}$, $\Lambda_T$\} 
determined from the fits to the lattice data using the generalized scheme are given in Tables~\ref{table: octet GEN parameters} and \ref{table: Dec GEN parameters}, respectively.

\begin{table}[H]
\begin{center}
\caption{Octet baryon fit parameters $M_B^{(0)}$, $C^{(1)}_{B\ell}$, $C^{(1)}_{Bs}$, and $\Lambda_B$ ($B=N$, $\Lambda$, $\Sigma$, $\Xi$) for the generalized baryon mass expansion scheme.}
\begin{tabular}{ l | c | c c c c }
\hline
& $B$ & $M_B^{(0)}$ & $C^{(1)}_{B\ell}$ & $C^{(1)}_{Bs}$ & $\Lambda_B$ \\
&     & (MeV) & (MeV$^{-1}$) & (MeV$^{-1}$) & (MeV) \\
\hline
\multirow{4}{*}{~{\bf MDLS}~} 
& ~$N$~ & \multirow{4}{*}{ ~$870(32)$~ } 
& ~$-2\,738(256)$~ & ~~~$-281(58)$~     & ~$678(118)$~
\\
& ~$\Lambda$~ & 
& ~$-2\,186(199)$~ & ~~~$-869(55)$~     & ~$680(110)$~
\\
& ~$\Sigma$~ & 
& ~$-1\,858(302)$~ & ~~$-1\,108(66)$~~ & ~$442(417)$~
\\
& ~$\Xi$~ &
& ~$-1\,226(229)$~ & ~\,$-1\,631(114)$   & ~$448(136)$~
\\
& & & &
\\
\multirow{4}{*}{~{\bf MILS}~}
& ~$N$~ & \multirow{4}{*}{~$927(26)$~} 
& ~$-2\,620(192)$~ & ~~~$-233(52)$~     & $730(92)$~
\\
& ~$\Lambda$~ &
& ~$-2\,108(153)$~ & ~~~$-782(44)$~     & ~\,$734(238)$~
\\
& ~$\Sigma$~ &
& ~$-1\,801(178)$~ & ~~~$-992(80)$~     & $387(69)$~
\\
& ~$\Xi$~ &
& ~$-1\,234(179)$~ & ~~$-1\,491(80)$~~  & ~\,$494(158)$~
\\
\hline
\end{tabular}
\label{table: octet GEN parameters}
\end{center}
\end{table}

%
\begin{table}[H]
\begin{center}
\caption{Decuplet baryon fit parameters $M_T^{(0)}$, $C^{(1)}_{T\ell}$, $C^{(1)}_{Ts}$, and $\Lambda_T$ ($T=\Delta$, $\Sigma^*$, $\Xi^*$, $\Omega$) for the generalized baryon mass expansion scheme.}
\begin{tabular}{ l | c | c c c c }
%
\hline
& $T$ &  $M_T^{(0)}$ & $C^{(1)}_{T\ell}$ &  $C^{(1)}_{T s}$ &  $\Lambda_T$ \\
&     & (MeV) & (MeV$^{-1}$) & (MeV$^{-1}$) & (MeV) \\
\hline

\multirow{4}{*}{\bf MDLS}  
& ~$\Delta$~  & \multirow{4}{*}{~$1\,184(39)$}
                  & ~$-$2\,014(544)~ & ~~~$-370(191)$~  & ~601(224)~ 
\\
& ~\,$\Sigma^*$ & & ~$-1\,478(547)$~ & ~~~$-851(202)$~  & ~508(459)~
\\
& ~\,$\Xi^*$    & & ~~$-898(455)$  &  ~~$-1\,377(145)$~~ & ~499(652)~
\\
& ~$\Omega$~    & & ~~$-407(272)$  &  ~~$-1\,861(160)$~~ & ~540(423)~
\\
& & &
\\
\multirow{4}{*}{\bf MILS} 
& ~$\Delta$~ & \multirow{4}{*}{~$1\,254(54)$}  
                  & ~$-2\,027(545)$~ & ~~~$-294(171)$~  & ~636(214)~  
\\
& ~\,$\Sigma^*$ & & ~$-1\,545(456)$~ & ~~~$-733(199)$~  & ~500(82)~~\,
\\
& ~\,$\Xi^*$    & & ~$-1\,002(400)$~ & ~~$-1\,238(185)$~~ & ~494(124)~
\\
& ~$\Omega$~    & & ~~$-541(251)$ & ~~$-1\,698(191)$~~ & ~582(511)~
\\
\hline
\end{tabular}
\label{table: Dec GEN parameters}
\end{center}
\end{table}

\newpage
\subsection{$\sigma$-term results}
\vspace*{-0.2cm}

The octet baryon $\sigma$-terms and masses obtained from the fitted parameters in Tables~\ref{table: WL parameter values}--\ref{table: Dec GEN parameters} for the SU(3) contrained and generalized mass expansion schemes, and for the MDLS and MILS scenarios, are listed in Table~\ref{tab.sigmaterms-oct}, along with the $\chi^2$ per degree of freedom ($\chi^2_{\rm dof}$) values. 
The corresponding decuplet baryon results are given in Table~\ref{tab.sigmaterms-dec}.

\begin{table}[H]
\begin{center}
\caption{Octet baryon $\sigma$-terms and masses from the SU(3) contrained and generalized schemes, for the MDLS and MILS scenarios, along with the corresponding $\chi^2_{\rm dof}$ values.}
\begin{tabular}{ r | c  c  c  c | c  c  c  c }
\hline
& \multicolumn{4}{ c |}{{\bf MDLS}}  & \multicolumn{4}{c}{{\bf MILS}}\\
~$B$~
& ~$\sigma_{B\ell}$~ & ~$\sigma_{Bs}$~ & ~$M_B$~ & ~$\chi^2_{\rm dof}$~~ 
& ~$\sigma_{B\ell}$~ & ~$\sigma_{Bs}$~ & ~$M_B$~ & ~$\chi^2_{\rm dof}$~~
\\
& ~(MeV)~ & ~(MeV)~ & ~(MeV)~ & 
& ~(MeV)~ & ~(MeV)~ & ~(MeV)~
\\
\hline
~{\bf SU(3) constrained}~~~~~~
&  &  &  &  &  &  &  &
\\
~$N$~ 
& ~41(4)~ & ~\,50(8) & ~~~\,910(14) & \multirow{4}{*}{~~0.89~~}
& ~46(3)~ & ~\,49(8)  & ~~~\,930(14) & \multirow{4}{*}{~~0.78~~}
\\
~$\Lambda$~
& 29(2) & 203(7) & 1\,070(8) & 
& 32(2) & 189(7) & 1\,089(8) &
\\
~$\Sigma$~
& 24(2) & 263(7) & 1\,132(7) & 
& 26(2) & 248(7) & 1\,158(7) &
\\
~$\Xi$~ 
& 14(1) & 376(7) & 1\,257(3) & 
& 15(1) & 353(6) & 1\,280(4) &
\\
& & & & & & & & 
\\
~{\bf generalized}\hspace*{1.7cm}
&  &  &  &  &  &  &  &
\\
~$N$~  
& ~46(5)~ & ~\,68(15) & ~~~896(18) & \multirow{4}{*}{~~0.82~~} 
& ~47(4)~ & ~\,59(14) & ~~~921(18) & \multirow{4}{*}{~~0.75~~}  
\\
~$\Lambda$~
& 32(7) & 212(12) & ~1\,055(12) & 
& 34(3) & 200(13) & ~1\,076(12) & 
\\
~$\Sigma$~
& 23(8) & 257(11) & ~1\,138(15) &
& 21(3) & 228(22) & ~1\,170(10) & 
\\
~$\Xi$~
& 14(1) & 379(16) & 1\,260(4)\, & 
& 14(1) & 348(17) & 1\,283(4)\, &
\\
\hline
\end{tabular}
\label{tab.sigmaterms-oct}
\end{center}
\end{table}

\begin{table}[H]
\begin{center}
\caption{Decuplet baryon $\sigma$-terms and masses from the SU(3) contrained and generalized schemes, for the MDLS and MILS scenarios, along with the corresponding $\chi^2_{\rm dof}$ values.}
\begin{tabular}{ r | c  c  c  c | c  c  c  c }
\hline
& \multicolumn{4}{ c |}{\bf MDLS } & \multicolumn{4}{c}{\bf MILS }\\
~$T$~  & ~$\sigma_{T\ell}$~ & $\sigma_{Ts}$ & $M_{T}$ & ~$\chi^2_{\rm dof}$~ & ~$\sigma_{T\ell}$~ & $\sigma_{Ts}$ & $M_{T}$ & $\chi^2_{\rm dof}$\\
& ~(MeV)~ & ~(MeV)~ & ~(MeV)~ & 
& ~(MeV)~ & ~(MeV)~ & ~(MeV)~
\\
\hline
~{\bf SU(3) constrained}~~~~~~
&  &  &  &  &  &  &  &
\\
~$\Delta$~  
&  ~~26(13)~ & ~\,69(17) & ~1\,240(38) 
            & \multirow{4}{*}{~~3.29~}
&  ~~31(11)~ & ~\,64(15) & ~1\,285(40) 
            & \multirow{4}{*}{~~2.70~}
\\
~\,$\Sigma^*$  
& ~16(9)\,~ &  198(18)   & ~1\,363(17) & 
& ~19(7)\,~ &  180(14)   & ~1\,407(20) & 
\\
~\,$\Xi^*$
&  ~8(5)     & 322(19)   & \,1\,499(9)~ &
&  ~11(4)\,~ &  292(16)  & \,1\,540(9)~ & 
\\
~$\Omega$~
& ~4(2)      &  437(19)  & ~1\,645(10) & 
& ~5(2)      &  398(20)  & ~1\,681(12) & 
\\
& & & & & & & & 
\\
~{\bf generalized}\hspace*{1.7cm}
&  &  &  &  &  &  &  &
\\
~$\Delta$~ 
& ~~38(17)~  & ~\,88(50)  & ~1\,209(71)  & \multirow{4}{*}{~~4.39~} 
& ~~40(15)~  & ~\,71(48) & ~1\,245(70)   & \multirow{4}{*}{~~3.61~} 
\\
~\,$\Sigma^*$ 
& ~23(23)    &  199(46)  & ~1\,368(55) & 
& ~24(8)\,~  &  171(46)  & ~1\,412(29) & 
\\
~\,$\Xi^*$
&  ~13(14)   &  321(39)  & ~1\,500(31) &
&  ~14(6)\,~ &  288(45)  & ~1\,539(23) &
\\
~$\Omega$~  
& ~4(3)      & 435(25)   & ~1\,622(11) &
& ~5(3)      & 400(38)   & ~1\,655(12) &  
\\
\hline
\end{tabular}
\label{tab.sigmaterms-dec}
\end{center}
\end{table}

\twocolumngrid
\section{Convergence of finite range regulated chiral effective theory}
\label{sec: Convergence}

Since the extrapolation of the lattice data is performed with SU(3) chiral effective theory, one may question the importance of higher order loop corrections in our calculation and the convergence of the effective theory.
Considering an expansion of $\sigma_{\pi N}$ in powers of the pion mass, we compare the relative sizes of terms at the physical point.
We find the ratio of the ${\cal O}(m_\pi^3)/{\cal O}(m_\pi^2)$ terms to be $\approx -0.35$, indicating that the ${\cal O}(m_\pi^3)$ contribution gives~a $\sim 30\%$ correction to the leading order result. 
Note that the ${\cal O}(m_\pi^3)$ contribution does not represent our $\delta \sigma_{\pi N}^{3/2}$ correction (analogous to the $\delta M_{\cal B}^{3/2}$ term in Eq.~(\ref{eq: chiral mass expan})), which estimates all higher order terms by virtue of the FRR fitting parameter. 
The ${\cal O}(m_\pi^3)$ term is the leading nonanalytic term predicted from QCD, and hence must be independent of the regularization scheme or the effective field theory implementation used. 

\begin{figure}[h]
\centering
\includegraphics[width=0.5\linewidth]{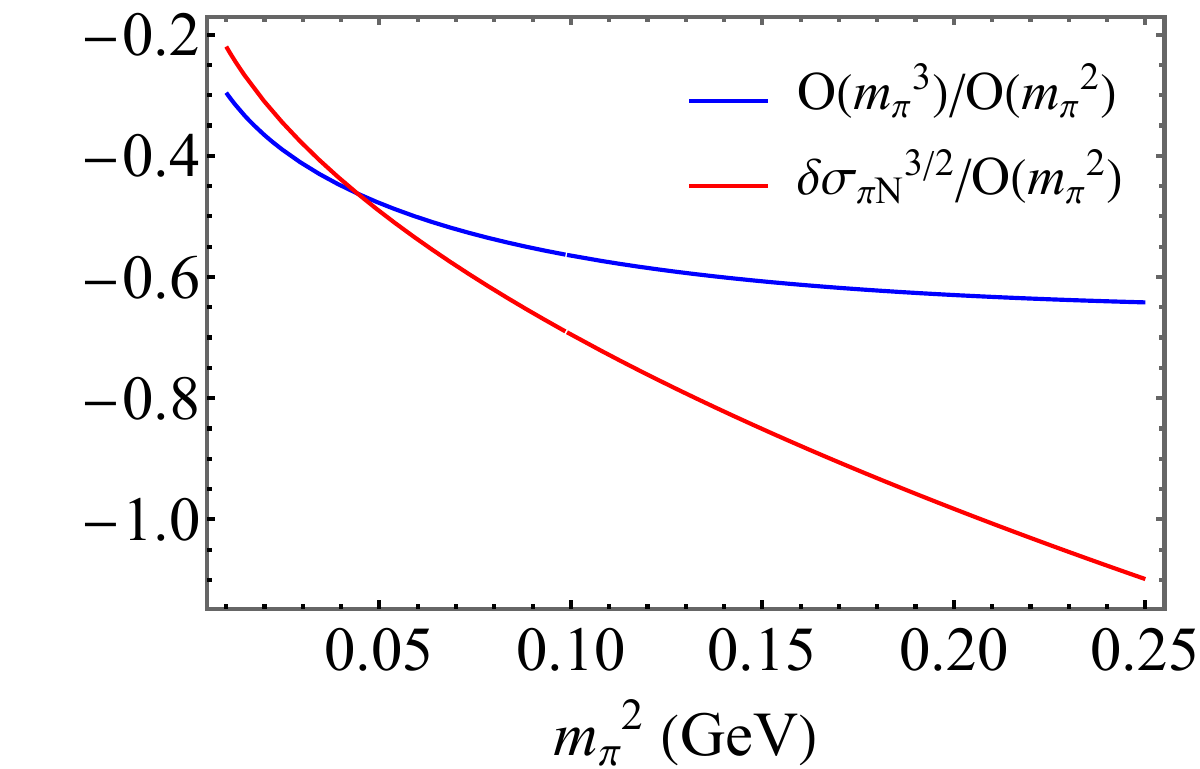}
\caption{Ratio of terms in the chiral expansion of $\sigma_{\pi N}$, relative to the ${\cal O}(m_\pi^2)$ term. The ratio ${\cal O}(m_\pi^3)/{\cal O}(m_\pi^2)$ (blue) for the LNA ${\cal O}(m_\pi^3)$ term is compared with the ratio $\delta \sigma^{3/2}_{\pi N}/{\cal O}(m_\pi^2)$ (red), where $\delta \sigma^{3/2}_{\pi N}$ contains higher order terms in powers of $m_\pi$ parameterized by the finite range regulators.} 
\label{fig: sigmaratiowlegend}
\end{figure}

At the physical point we find the ratio $\delta \sigma_{\pi N}^{3/2}/{\cal O}(m_\pi^2)$ to be $\approx -0.36$, which is comparable to the ${\cal O}(m_\pi^3)/{\cal O}(m_\pi^2)$ ratio, suggesting the net effect of the ${\cal O}(m_\pi^4)$ and higher terms captured by the FRR prescription is relatively small here. 
At larger pion masses the ratio $\delta \sigma_{\pi N}^{3/2}/{\cal O}(m_\pi^2)$ varies dramatically from the ${\cal O}(m_\pi^3)/{\cal O}(m_\pi^2)$ ratio, as seen in Fig.~\ref{fig: sigmaratiowlegend}.
This suggests that the higher order corrections become more important at larger $m_\pi^2$, as expected.
These corrections are well described by our finite range regulator, as indicated by the quality of the fit to the data in Fig.~\ref{fig: Combined Hybrid fits}.

\onecolumngrid

\twocolumngrid

\end{document}